# ANALYSIS OF POLYSILICON MICRO BEAMS BUCKLING WITH TEMPERATURE-DEPENDENT PROPERTIES

*M. Shamshirsaz[1], M. Bahrami[2], M. B. ASGARI[3] and M. Tayefeh[2]*

1. New Technologies Research Center (NTRC), 2. Mechanical Engineering Department
Amirkabir University of Technology (Tehran Polytechnic)
3. Energy Engineering Department, Power and Water University of Technology (PWUT), Tehran, Iran
E-mail: shamshir@aut.ac.ir

**ABSTRACT**

The suspended electrothermal polysilicon micro beams generate displacements and forces by thermal buckling effects. In the previous electro-thermal and thermo-elastic models of suspended polysilicon micro beams, the thermo-mechanical properties of polysilicon have been considered constant over a wide rang of temperature (20-900°C). In reality, the thermo-mechanical properties of polysilicon depend on temperature and change significantly at high temperatures. This paper describes the development and validation of theoretical and Finite Element Model (FEM) including the temperature dependencies of polysilicon properties such as thermal expansion coefficient and Young's modulus. In the theoretical models, two parts of elastic deflection model and thermal elastic model of micro beams buckling have been established and simulated. Also, temperature dependent buckling of polysilicon micro beam under high temperature has been modeled by Finite Element Analysis (FEA). Analytical results and numerical results using FEA are compared with experimental data available in literature. Their reasonable agreement validates analytical model and FEM. This validation indicates the importance of including temperature dependencies of polysilicon thermo-mechanical properties such as Coefficient of Thermal Expansion (CTE) in the previous models.

## 1. INTRODUCTION

Electro-thermal actuation is an attractive choice in MEMS applications comparing with the other types of actuation based on electrostatic, magnetic, piezo-electric effects. This is due to their ease of fabrication and also their ability to generate large displacements and forces for micro-assembly and micro-positioning purposes. These suspended micro beams are made of heavily phosphorus-doped polysilicon and fabricated by surface micromachining technology using the Multi-User MEMS Process (MUMPS) [1,2].

These suspended micro actuators operate by deformation due to buckling phenomena produced by localized thermal stresses. When an electrical current pass through this clamped-clamped micro beams, thermal expansion caused by joule heating generates thermal stresses, which are in origin of buckling. The performance of these micro actuators is strongly influenced by thermo-mechanical behavior of material during buckling phenomena. Therefore, in order to design these micro actuators, it is essential to characterize thermo-mechanical behavior of micro actuators and consequently to determine corresponding temperature for desired deflection.

Many works have been done in modeling electro-thermal and thermo-mechanical behavior of thermal flexure actuators [3-13]. R.S.Chen *et al.* used FEA to model an electro thermal micro actuator and evaluate its tip deflection and assumed constant thermo-mechanical properties over a wide rang of temperature [3]. The thermal behavior of thermo-mechanical in-plan micro actuators has been modeled using a finite difference approach by Lott *et al.* [4]. The variation of polysilicon thermal conductivity and resistivity with temperature has been considered in this model. The thermal behavior of thermo-mechanical flexure microactuators has been analyzed analytically and numerically using FEA by Mankame and Ananthasuresh [5]. Thermal conductivity and Coefficient of Thermal Expansion (CTE) are taken into account as a function of temperature. A thermal model have been presented for U-beam and V-beam thermal actuators considering the electrical and thermal conductivity polysilicon variations with temperature by Geisberger *et al.* [11].Also, the doping variation influences on electrical and thermal conductivity within polysilicon and single crystal silicon have been explored [11]. A three-dimensional non-linear FEM has been developed for an out-of-plan thermal actuator considering thermo-physical property variations [12]. The temperature dependent behavior of thermal flexure micoactuators has been analyzed by Shamshirsaz and Gheisarieha [13]. It has been shown that the effect of material property temperature dependencies becomes important in high currents.

Few works have been done in modeling electro-thermal and thermo-mechanical behavior of suspended thermal micro actuators. The electro-thermal and elastic characterizations of suspended micro beams have been investigated by Lin and Chiao [14-16]. The coupled electro-thermal and elastic models of thermally driven clamped-clamped micro beams have been analyzed by M. Chiao and L. Lin [15]. They observed and reported a significant deviation of results obtained by theoretical models from experimental results [14]. They have already characterized the sensitivity of these micro beams with respect to different process variations such as: effect of geometrical variations due to the undesired manufacturing process changes (a trapezoidal section instead of a square shape), in fact, width and thickness of beams vary from different wafers or different locations of square wafer. They reported that these geometrical variations can generates a deviation of analytical results from experimental results about 10% [14]. Also, M. Chiao and L. Lin discussed about the deviation of theoretical and FEA results from experimental results. They informed that the variations in material properties with respect to temperature have not been considered in their models. They suggested a detail study of thermal and mechanical properties changes with temperature is essential to improve the accuracy of theirs results.

In reality, the variation of Young's modulus and thermal expansion coefficient are significant with temperature [17, 18]. By new techniques and procedures, recently, the tensile tests of polysilicon have been realized and the results indicated a decrease of Young's modulus with temperature rise [17]. Also, thermal expansion coefficient increase with temperature rise during micro actuator operation. The variation of thermal expansion coefficient for polysilicon is reported by C.H. Pan [18]. This report shows that polysilicon thermal expansion coefficient varies significantly over the temperature range of 450 °C and higher. No significant change in Poisson's ratio and fracture strength with temperature is reported [17, 18].

It is, therefore, necessary to consider the nonlinearities introduced by temperature-dependent properties when the buckling behavior of an electro-thermally driven micro actuator is to be accurately modeled over a wide rang of temperature.

The time-dependent buckling of suspended polysilicon micro beams with temperature-dependent properties has been analyzed previously using FEA by Shamshirsaz *et al.* [19].

In this study, the temperature-dependent buckling of suspended polysilicon micro beams has been investigated analytically and numerically using FEA. The analytical and FEM results are compared with experimental results obtained by M. Chiao and L. Lin [14].

In the first step of this study, previous analytical models have been modified in order to introduce the temperature dependency of polysilicon properties. The thermo-elastic model for these micro actuators has been established by coupled problems of Elastica and Duhamel-Neumann law previously [20]. A computer simulation program has been developed in MATLAB environment to determine the different parameters in thermo-mechanical equations.

In the second step of this study, temperature-dependent buckling of polysilicon micro beam under high stress with temperature-dependent material properties has been modeled and studied by FEA. Using FEA, the relative importance of temperature dependencies for various parameters can be examined. A one-dimensional finite element model of clamped–clamped polysilicon micro beam is developed.

The goal is to provide valid theoretical and numerical models for buckling phenomena of micro beams considering temperature-dependent material properties.

## 2. TEMPERATURE DEPENDENT MODEL

A diagram of a deformed clamped-clamped polysilicon micro beam is shown in figure 1. It generates deflection through thermal loads P, due to temperature rise.

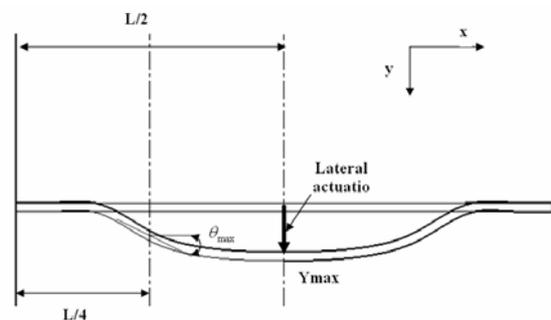

**Figure 1.** Deflection of a clamped-clamped micro beam under thermal loads.

Due to symmetry, only one-quarter of micro beam is analyzed (figure 2). Analytical models are developed in two parts: elastic and thermal elastic deflection based on Lin and chiao theoretical models [15]. Then, the temperature dependency of parameters is integrated in these models. By these models, the thermal load and maximum deflection relations are derived as:

$$T = \frac{4K(\beta)E_T A\sqrt{PE_T I} - PLE_T A - P^2 L}{PLE_T A\alpha_T} + T_0$$

$$\gamma_{max} = 4\beta\sqrt{\frac{I}{[-\varepsilon + \alpha_T(T-T_0)]A}}$$

with $\beta = \sin\frac{\theta_{max}}{2}$ and $\varepsilon = \frac{L'-L}{L}$

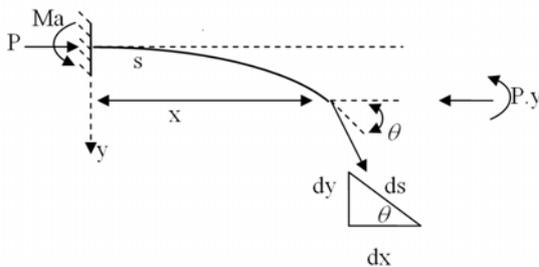

**Figure 2.** One-quarter of deflected micro beam

Where

P: thermal loads in axial direction
$M_a$: unknown reaction moment at clamped edge
$\theta$ : deflection angle
s: coordinate along the deflected micro beam
$\gamma_{max}$ : maximum deflection
I: moment of inertia
$E_T$: Young's modulus relation with temperature
$\alpha_T$ : thermal expansion coefficient relation with temperature
L: initial length beam
$L'$ : deflected length beam
A: cross-section area of micro beam
T: average temperature of the beam
$T_0$: reference temperature
$\alpha$ : thermal expansion coefficient
$K(\beta)$ : complete elliptic integral of the first kind

The initial values of Young's modulus, Poisson ratio and thermal expansion coefficient are considered: 150e+9 Pa, 0.22, 21.6e-6 °C$^{-1}$ respectively in ambient temperature. The data related to the variation of Young's modulus and thermal expansion coefficient (TEC) reported by Sharp *et al* [17] and Pan [18] is applied to the theoretical and numerical models (see Appendix).

### 3. ANALYTICAL SIMULATION RESULTS

In this study, elastic deflection model and thermal elastic model of clamped-clamped micro beam have been developed using computer simulation in MATLAB environment. A micro beam with 1 μm width, 1.5 μm thickness and 100 μm length is used as simulation sample. A first estimation of buckling temperature without considering temperature dependency of properties is determined. An average temperature is supposed in beam length. In the next steps, a program loop is established to determine the appropriate properties in each corresponding temperature. Generally, all of the related temperatures for buckling and post-buckling stage obtained by the models in this study are lower than those obtained by previous models (fig. 3). The experimental results obtained for maximum deflection versus current have manifested a shift to the left side comparing with theoretical results obtained from previous models, which did not take to account the temperature dependency of polysilicon properties. This confirms that these results are closer to experimental results.

### 4. FINITE ELEMENT SIMULATION RESULTS

A one-dimensional model of clamped-clamped micro beam is made using FEA simulation program. A model similar to simulation sample in analytical model has been considered. This model has 100μm length, 1 μm wide and 1.5 μm thick.

A steady state structural analysis with coupled thermo-mechanical behavior is performed. A non-linear model with large deformation is developed for post-buckling state. Shear deflection is considered, as well. In order to take account the variations of Young' modulus and thermal expansion coefficient with temperature, the temperature is incremented in small steps and the corresponding values of these parameters are used in the analysis. The effect of imperfections such as fabrication defects, geometric irregularities and non-ideal loads on the structure deformation is similar, in the post-buckling regime. Therefore, as an approximation, the imperfection effect is modeled by a pre-deformation of micro beam in its natural unstressed state. To introduce this pre-deformation, an out-of-plan load proportional to micro beam length was applied to this model. The temperature of clamped ends is considered ambient temperature at all times. Finite element analysis provides us with steady state and current-temperature response, deflections measured at middle of micro beam length.

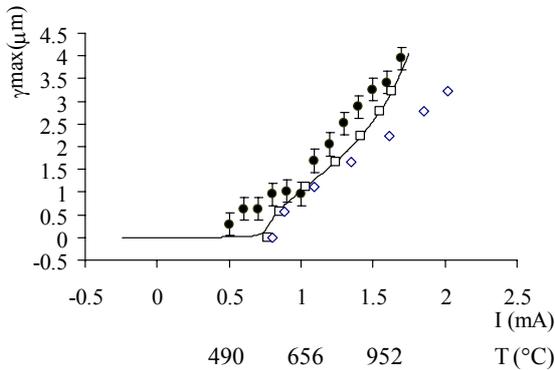

**Figure 3.** Maximum Deflections $\gamma_{max}$ versus Current / temperature:
● experimental, □ analytical and **-** FEM with temperature dependent properties, ◊ analytical without temperature dependent properties results.

## 5. DISCUSSION

When the dependency of the material properties on temperature is taken into account in the model, the deflection-temperature curve shows a shift to the left; i.e., buckling happens in a lower temperature. We can also observe from 'figure 3' that, without considering the temperature dependency, buckling deflection is underestimated and buckling temperature is overestimated compared to experimental results.

The theoretical results obtained without considering the temperature dependency of material properties show an important deviation from experimental results. The related Root Mean Square (RMS) is about 3.2 μm. Considering the temperature dependency of properties; the analytical results curve has a shift to left side and become closer to experimental results. The corresponding RMS becomes about 1.9 μm.

Applying the temperature dependency of material properties to FEM, the related results coincide with analytical results obtained by temperature dependent model (RMS = 1.8μm) except in initiating buckling regime. This is because, an out of plan load representing the imperfections in pre-buckling regime is considered in FEM, consequently, the growth of micro beam deflection becomes smooth and continuous. Therefore, in place of the buckling critical point, a region of transition from small to large deflections or to buckling region is observed in 'figure 3'.

As we observe in 'figure 3', the difference between experimental results and theoretical results obtained by previous model increases with temperature. This can be due to the gradient increase of TEC with temperature (figure 4-Appendix). This effect show an evident predominance of deviation produced by considering a constant TEC over temperature operational rang of micro beam. So, applying previous models, the prediction of micro beam deflection is effectively impossible especially in high temperature regime generated by high current in micro beam. The proposed models in this study improve analytical and FEM results. By these observations, it is found that the temperature-dependent simulation results not only get closer to experimental results presented in [14] but also, the deviation of these results from experimental results stay almost constant for the different temperature regimes.

Although, the results obtained by the model proposed in this study are effectively closer to experimental results, but there still exist a deviation by an underestimation of micro beam deflections. This can be probably due to manufacturing process variations. The effect of dimension variations by manufacturing process has been examined by M. Chiao and L. Lin [14]. They reported that this latter can generates a variation of analytical results about 10%.

## 6. CONCLUSIONS

The simulation results without considering the temperature dependency of properties underestimate the beam lateral deflections. It has previously been observed a difference between experimental results and analytical results. In this study, the simulation results confirm major deviation from experimental results is due to the variation of TEC of polysilicon with temperature. It is also shown that by introducing the variation of polysilicon properties with temperature in analytical models, we can effectively correct this difference.

Finite element nonlinear analysis is developed for polysilicon micro beam considering temperature dependency of material properties. This shows that the previous works, which did not consider the temperature dependency of polysilicon properties, underestimate the buckling critical load and the deflection. The results also show that the deviation produced is rather considerable. The agreement between FEA results and experimental results suggests that, the temperature dependencies of polysilicon properties and transient response of micro beam should be considered for prediction of post-buckling behavior of polysilicon micro beam.

# 8. APPENDIX

Young's modulus temperature dependency [17]:

$E(T) = E_s - 0.04(T - T_S)$

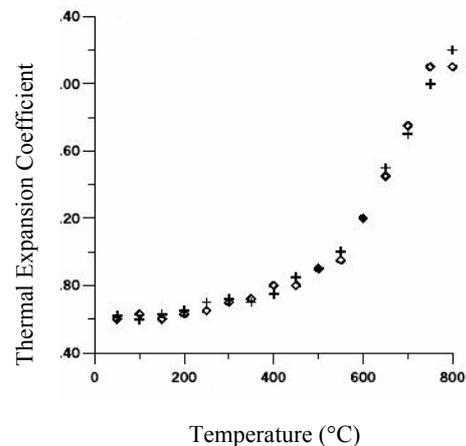

**Figure 4**. Thermal expansion coefficient versus temperature [18].